 \documentclass[final,5p,times,twocolumn,authoryear]{elsarticle}

\usepackage{amssymb}
\usepackage{lipsum}
\usepackage{graphicx}
\usepackage{txfonts}
\usepackage{hyperref}
\usepackage{color}
\usepackage{url}
\usepackage{comment}
\usepackage{subcaption}

\usepackage{float}    
\usepackage{listings}
\usepackage{xcolor}

\journal{Astronomy $\&$ Computing}

\begin{document}

\begin{frontmatter}



\title{SAS in ESA Datalabs: A New Platform for XMM-Newton Analysis}



\author[affl:fau]{Esin G. Gülbahar}
\ead{esin.gulbahar@fau.de}
\author[affl:esa]{Camille M. Diez}
\author[affl:tpz]{Aitor Ibarra}
\author[affl:tpz]{Ivan Valtchanov}
\author[affl:tpz]{Richard Saxton}
\author[affl:quasar]{Ignacio de la Calle P\'erez}
\author[affl:aurora]{Jose Lopez-Miralles}
\author[affl:CIEMAT]{Alejandro Gonz\'alez Ganzábal}
\author[affl:esa]{Peter Kretschmar}
\affiliation[affl:fau]{organization={FAU Erlangen-Nürnberg},
            addressline={Dr. Karl Remeis-Sternwarte, Sternwartstr. 7}, 
            city={Bamberg},
            postcode={96049},
            country={Germany}}
\affiliation[affl:esa]{organization={European Space Agency (ESA)},
            addressline={European Space Astronomy Centre (ESAC), Camino Bajo del Castillo s/n}, 
            city={Villanueva de la Cañada},
            postcode={28692}, 
            state={Madrid},
            country={Spain}}
\affiliation[affl:tpz]{organization={Telespazio U.K. Ltd. for the European Space Agency (ESA)},
            addressline={European Space Astronomy Centre (ESAC), Camino Bajo del Castillo s/n}, 
            city={Villanueva de la Cañada},
            postcode={28692}, 
            state={Madrid},
            country={Spain}}
\affiliation[affl:quasar]{organization={Quasar Science Resources S.L. for the European Space Agency (ESA)},
            addressline={European Space Astronomy Centre (ESAC), Camino Bajo del Castillo s/n}, 
            city={Villanueva de la Cañada},
            postcode={28692}, 
            state={Madrid},
            country={Spain}}
\affiliation[affl:aurora]{organization={Aurora Technology for the European Space Agency (ESA)},
            addressline={European Space Astronomy Centre (ESAC), Camino Bajo del Castillo s/n}, 
            city={Villanueva de la Cañada},
            postcode={28692}, 
            state={Madrid},
            country={Spain}}
\affiliation[affl:CIEMAT]{organization={Centro de Investigaciones Energéticas, Medioambientales y Tecnológicas (CIEMAT)},
            addressline={Av. Complutense 40}, 
            city={Moncloa-Aravaca},
            postcode={28040}, 
            state={Madrid},
            country={Spain}}

\begin{abstract}
XMM-Newton is a cornerstone mission of the European Space Agency (ESA) for X-ray astronomy, providing high-quality X-ray data for astrophysical research since the start of the century. Its Science Analysis System (SAS) has been a reliable data reduction and analysis software, evolving throughout the years to meet changing user needs, while incorporating new methods. This paper presents the XMM-SAS Datalab, a tool within the cloud-based ESA Datalabs platform, designed to enhance the interactivity and collaborative potential of SAS. By integrating SAS with a modern, Python-based JupyterLab interface, it enables shared analysis workspaces, removes the need for local software setup, and provides faster access through containerised environments and preconfigured libraries. Moving SAS to the cloud preserves a consistent software setup while eliminating installation complexities, saving time and effort. A case study of the X-ray binary Vela X-1 demonstrates that the Datalabs platform reliably replicates local SAS outputs, with minimal deviations attributed to calibration file versions. The XMM-SAS Datalab allows straightforward X-ray data analysis with collaborative process, setting the way for future adaptations in e-science platforms and multi-wavelength astronomy, while offering traceability and reproducibility of scientific results.
\end{abstract}




\end{frontmatter}




\section{Introduction}
\label{introduction}

XMM-Newton is an X-ray observatory launched by the European Space Agency (ESA) in 1999 onboard Ariane-5 as part of its Horizon 2000 program \citep{Lumb:2000, Jansen:2001}. It has significantly contributed to our understanding of the X-ray universe by providing high-quality spectral and imaging data. The primary objective of XMM-Newton was to produce high-quality X-ray spectroscopy of faint sources, as a follow-up on the predecessor X-ray mission EXOSAT \citep{EXOSAT}. To achieve its objectives, XMM-Newton is equipped with telescopes that have large collecting areas. The design includes three multi-mirror telescopes, each consisting of 58 Wolter I mirrors \citep{Aschenbach_1986}. The instrumentation onboard the observatory consists of Three European Photon Imaging Cameras (EPIC), where two of them are MOS-CCD \citep{Turner_2001} and one of them being the pn-CCD cameras \citep{Struder_2001}, two Reflection Grating Spectrometers (RGS) \citep{den_Herder_2001}, and one Optical Monitor (OM) \citep{Mason_2001}.

Developed by the XMM-Newton Science Operations Centre and the Survey Science Center, the XMM-Newton Science Analysis System (SAS) is a software built to reduce and calibrate the data collected by the XMM-Newton observatory \citep{Gabriel_2004, Gabriel:2017}. SAS is a necessary tool for any researcher who wants to process XMM-Newton data. Furthermore, to ensure correct scientific analysis, the user will have to utilise the data set Current Calibration Files (CCF) \citep{Jansen:2001}. Setting up the skeleton of this analysis pipeline takes time and space. Previous efforts have been made to work around the SAS installation process by providing the XMM-Newton Remote Interface to Science Analysis (RISA) software, made available in link with the XMM-Newton Science Archive (XSA) \citep{RISA, RISA2}. RISA was developed to allow the user to do on-the-fly data analysis, through an interactive web interface. Whilst providing a quick solution to setup, RISA does not allow for continuity of work in one platform, or the ability to build personal scripts, restricting in-depth customized workflows. 

Providing compact, modern, and streamlined software for data reduction and analysis is a common concern among various fields of Astronomy. Cloud environments offered through various science platforms not only provide an alternative to this concern, but also ensure a collaborative environment by allowing to share common workspaces. Due to its capabilities, SAS requires a complex setup, posing challenges for new users. To address these challenges, we introduce the XMM-SAS Datalab, a new environment within the European Space Agency's \href{https://datalabs.esa.int/}{ESA Datalabs} \footnote{\url{https://datalabs.esa.int/}}, an e-science platform providing access to ESA archives and pre-set virtual environments. Users can run their code via a web interface from any device, seamlessly accessing SAS analysis through an online JupyterLab interface, preserving the interactive elements of the local SAS build.

It provides SAS through a modern lens, with comfortable interaction through the Python programming language and collaborative workspaces for team/class members. In accordance with the capabilities of the platform, the XMM-Newton researcher can create their own private or public environments by building their own Docker images in accordance with their research requirements. 

In this paper, we aim to describe what ESA Datalabs offers to SAS usage by investigating the principle features of the platform and the advantages of using SAS in the cloud environment. Section \ref{sec:XMM-Newton Science Analysis System} gives a quick introduction to SAS. Here, we also explore pySAS, a Python wrapper for SAS commands that allows users to execute tasks in a Python environment. Section \ref{sec:The Datalabs Platform} presents a general overview of the ESA Datalabs platform. Here, we outline the main functionalities and the products that the platform has to offer the user. The implementation of SAS is described in Section \ref{sec:XMM-SAS as a Datalab} while demonstrating the available features and methods implemented to keep the interactivity of the local SAS. To demonstrate the capability of the platform, we provide a case study of the well-studied X-ray binary Vela X-1, by replicating a previous analysis done by \cite{Diez_2023} in Section \ref{sec:case_study}. In Section \ref{sec:discussion}, we discuss the benefits and trade-offs of ESA Datalabs for SAS while investigating other services and discussing the implications of the scientific case study of Vela X-1. Finally, in Section \ref{sec:conclusion} we summarise the points discussed so far and present our overall findings.

\section{XMM-Newton Science Analysis System} \label{sec:XMM-Newton Science Analysis System}
For over two decades, SAS has provided researchers with the tools to process raw data files and extract scientific products of interest. Continuously expanding its capabilities, it offers a wide range of tasks and algorithms. For more details, the reader is advised to consult the XMM-Newton SAS \href{https://xmm-tools.cosmos.esa.int/external/xmm_user_support/documentation/sas_usg/USG/}{User Guide} \footnote{\url{https://xmm-tools.cosmos.esa.int/external/xmm_user_support/documentation/sas_usg/USG/}}. 

\subsection{General capabilities and Setup}
The primary objective of SAS is producing calibrated event lists, images, spectra, detector response matrices, and source lists \citep{Gabriel_2004}. SAS has been developed in C++, Fortran 90/95, Python, and Perl, whilst also utilising open software such as ds9, Grace, HEASOFT, cfitsio, pgplot, fftw, and Qt \citep{Gabriel:2017}. The user interacts by running tasks that include different parameters from the command line and setting the mandatory parameters. SAS also has interactive features for running tasks, having incorporated a GUI system, where the user can access the input parameters for the tasks on the command line if necessary \citep{Gabriel_2004}. The GUIs make the tasks easy to visualise with their parameters. Furthermore, the user can utilise the interactive imaging and data visualisation application ds9 \citep{ds9} for region selections, which easily communicates with SAS. After each data reduction session, the user can access the log file which will contain information on all the tasks ran with the defined parameters, regardless of whether the tasks were performed from the command line or GUI.

SAS installation options can be found at the \href{https://www.cosmos.esa.int/web/xmm-newton/sas-download}{XMM-Newton Cosmos ESA website} \footnote{\url{https://www.cosmos.esa.int/web/xmm-newton/sas-download}} available for different platforms and operating systems. Currently, at the time of writing, the most up-to-date SAS version is v21.0. Please refer to the Appendix (see sec. \ref{setup appendix}) for further details. There are also options for a Virtual Machine for VMWare and a SAS Docker image based on the Linux Ubuntu 22.04LTS. These two options might be useful in avoiding extensive software set-up, as it will come pre-configured within the container environment. It is mainly beneficial if you have any other operating systems than the supported systems and would like to use SAS.

However, SAS comes with software requirements before setting up. For version 21.0 these include X11, Python, Perl, ds9, Grace, Heasoft, and WCSTools. As some of the tasks are Perl scripts, the user needs to install a specific version of Perl and set up required modules within the environment (see table \ref{tab:sas_software}). Since the user is also required to install all the XMM-Newton Current Calibration Files (CCF), these multiple stages of the installation and set-up process can pose challenges when trying to satisfy all requirements and takes time.

The SAS Docker image simplifies installation by packaging the SAS build system and its dependencies into a self-contained environment \citep{Marcos:2023}. The Docker allows all the build requirements, including specific software versions, to be provided in an isolated system, avoiding compatibility issues. This alternative allows for a practical approach to building SAS, with faster and more direct set-up, however, does not fully eliminate the necessity of installation. Furthermore, one should always consider that this does not address issues such as lack of ready-built SAS distributions, challenges in collaboration, or accessibility.

\subsection{pySAS} \label{pySAS}
SAS tasks can be executed within a Python or Jupyter environment through pySAS, a Python wrapper applied around SAS commands. This removes the gap between traditional command-line tools and popular scripting environments, allowing the creation of workflows integrated with Python libraries. The wrapper class can be imported directly from the pySAS main library. Objects from the wrapper class are defined by the task name and the input arguments of the task are passed as a list. Wrapper objects use the method \texttt{run()}, which is used to execute the task directly from the Python interface/Jupyter Notebook as if they were executed from the command line. This allows the user to create more complex scripts, fully compatible with current Python standards. Several examples of how pySAS is integrated into a Jupyter Notebook can be found in the XMM-Newton Data analysis threads \footnote{\url{https://www.cosmos.esa.int/web/xmm-newton/sas-threads}}.

Future tasks and utilities written in Python developed for SAS will also be imported as modules directly from the pySAS package. This leads to a more organic approach to import of future utilities. An example of this is the pyutils module (an AstroPy wrapper that mimics some functions from HEASARC’s FTOOLs for fits file manipulation \citep{FTOOLS_orig_paper}), which is imported directly as a submodule of pySAS.

\section{The ESA Datalabs Platform} \label{sec:The Datalabs Platform}
ESA Datalabs is a web-based cloud platform for scientific analysis. Datalabs provides access to ESA infrastructure and data archives online, where the user can actively develop their code \footnote{\url{https://datalabs.esa.int/help/Help.html}} \citep{Navarro2024}. There are multiple computational environments present such as established and new analysis tools, and older software, available through a web browser. The majority of the environments are run through a JupyterLab interface and customised for the needs of a specific mission or scientific domain. However, Datalabs also offers access to other tools such as Octave, Aladin, gLab, and TopCat, whilst maintaining a design that allows the addition of different tools. Access to ESA Datalabs is straightforward; it requires an \href{https://www.cosmos.esa.int/}{ESA Cosmos} \footnote{\url{https://www.cosmos.esa.int/}} account followed by a self-registration online form. Currently, ESA Datalabs is being actively developed and is provided in a beta version.

\subsection{Datalabs \& Workspaces}
ESA Datalabs provides the user with several applications or software tools that are purpose-built for scientific analysis. Some of them are environments with pre-built specific analysis software for specific ESA missions, whilst others are commonly used external tools such as DS-9. Each of these environment instances is called a 'Datalab'. 
 Only two instances of a datalab can be run simultaneously, however the user can make requests to the ESA Datalabs service desk for more. The platform provides a permanent and a temporary space for the user, where depending on individual needs, the user can request their space to be increased. Any file saved outside of the permanent \texttt{my\_workspace} or \texttt{team\_workspace} folders will be reset with the deletion of the datalab instance. For further safekeeping, the user can also connect to their git repositories within the datalabs and push their files of importance there. ESA Datalabs allows collaborative work within each datalab instance through the creation of team workspaces and folders.

\subsection{Data Volumes}\label{sec:datavolumes}
The platform provides access to datasets from various ESA missions and allows users to upload their data through the section \textit{Data Volumes}. There, the user can add a custom data volume, or by choosing the \textit{Add from catalog} option can browse those that ESA Datalabs offers. The volumes that have been added by the user get automatically mounted to any of the actively running datalab instances. The path to the mounted data volume is consequently determined to be in the \texttt{data} folder.

\subsection{Datalabs Editor: develop your own Datalabs} \label{sciapps}
The user can develop their custom datalab through the Datalabs Editor GUI. This tool allows searching, launching, and commenting on public datalabs. Furthermore, this is the place where users can develop and share, either privately or publicly, their custom-built datalabs, whilst also being able to test, delete, archive, or import. 

A custom datalab can be created in one of two allowed formats: a JupyterLab \footnote{\url{https://jupyterlab.readthedocs.io/en/stable/getting_started/overview.html}} or a VNC (Virtual Network Computing) based X server \footnote{\url{https://kasmweb.com/kasmvnc/docs/1.0.0/man/Xvnc.html}}. There are three main steps required to create a datalab. Initially, the user has to provide software and upload the \texttt{requirements.txt} file, then provide metadata, and finally build. If desired, users can base their datalab on an existing one. For example, if the researcher wants to build additional tools or software on top of an SAS datalab, they can use it as a starting image for their container.

For users interested in creating custom datalab instances or adding new plugins, detailed documentation is available in the \href{https://datalabs.esa.int/help/Datalabs_Editor.html}{Help section} of ESA Datalabs \footnote{\url{https://datalabs.esa.int/help/Datalabs_Editor.html}}, outlining the steps for software setup, metadata entry, and container deployment.

\section{XMM-SAS as a Datalab} \label{sec:XMM-SAS as a Datalab}
Within the ESA Datalabs environment, one of the pre-configured instances is the XMM-SAS Datalab (currently present with the name \textit{XMM-SAS21.0} following the latest available version). The user can run the built Docker container with the latest version of SAS available. The Dockerised SAS is a crucial part of the adaptation of SAS into e-science platforms. 

\begin{figure}[ht]
    \centering

        \centering
        \includegraphics[width=\hsize]{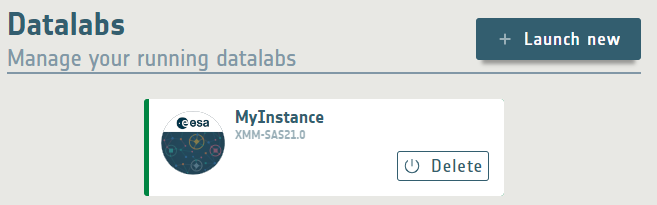}
        \caption{Selected instance of an XMM-SAS v21.0 Datalab, ready to be launched with a click. Using the \textit{Launch new} button, the user can view the catalgue of built datalabs and start a new instance.}
        \label{fig:sas_datalab}
   
\end{figure}

The Docker file, developed by the XMM-Newton Software Engineers, is based on JupyterLab, followed by the installation of HEASoft and then SAS. In the end, Python requirements necessary for a smoother analysis experience are added in the Docker container. A core feature of XMM-SAS Datalab is its use of pySAS, a Python wrapper that allows users to run SAS tasks directly from the JupyterLab interface of the container (see sec.\ref{pySAS}). The JupyterLab environment includes also some pre-built packages and tools for XMM-Newton data analysis (e.g. PyXspec \citep{pyxspec}). The user can find \textit{XMM-SASx.y} (where \textit{x.y} stands for the latest available SAS version) at the ESA Datalabs catalogue, and run the software by clicking on it (see Fig.\ref{fig:sas_datalab}). 

Using Python is not mandatory, as SAS commands can be called directly using the command prompt within JupyterLab. However, if the user wants to work directly within a Python environment, the pySAS wrapper can be imported to call any SAS task in the Jupyter Notebook:

\begin{lstlisting}
from pysas.wrapper import Wrapper as w
\end{lstlisting}

If the user would like to build on top of the existing environment, they can also create their environments or Datalabs (see sec.\ref{sciapps}). 
Therefore, the user does not need to configure the software. 
The shared team workspace feature allows SAS to be used in collaboration to foster a more enhanced research or learning experience. The collaborative space of Datalabs could become an environment for multi-wavelength research, where the user can easily switch between the SAS and other mission-specific datalabs. The updates of the XMM-SAS Datalab will be followed up by the XMM-Newton team to ensure access to the newest version.

\subsection{Available features}

\begin{figure}
   \centering
   \includegraphics[width=\hsize]{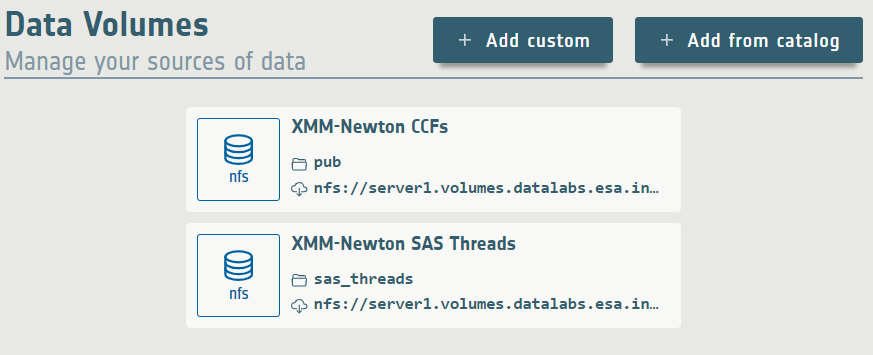}
      \caption{XMM-Newton CCFs and SAS threads can be mounted from the Data Volumes section of ESA Datalabs, available on the cloud. The folder names of each data volume is present on the respective boxes.}
         \label{fig:xmm data volume}
   \end{figure}

The XMM-Newton Current Calibration Files (CCF) are available in the Data Volume section of ESA Datalabs (see sec.\ref{sec:datavolumes}). There is direct access to the CCFs provided by the ESA Data Archive (see Fig.\ref{fig:xmm data volume}). When running the XMM-SAS Datalab, the data volume is automatically mounted and user interaction is not required, though they are available to be attached manually. The XMM-Newton Python SAS Threads for Datalabs (see sec.\ref{sec:threads}) are also available as a data volume, which is automatically mounted to each launch of XMM-SAS Datalab (see Fig.\ref{fig:xmm data volume}).  

The environment comes with pre-configured Python libraries to assist with the analysis of XMM-Newton data. For instance, for X-ray spectral fitting, \texttt{PyXspec} \citep{pyxspec}, the Python interface for Xspec, is built into the environment. A set of widely used tools are directly available for convenience, allowing uninterrupted, continuous research.

\subsubsection{Preserving interactivity}
\begin{figure}[ht]
    \centering
    \begin{subfigure}[a]{\hsize}
        \centering
        \includegraphics[width=5cm]{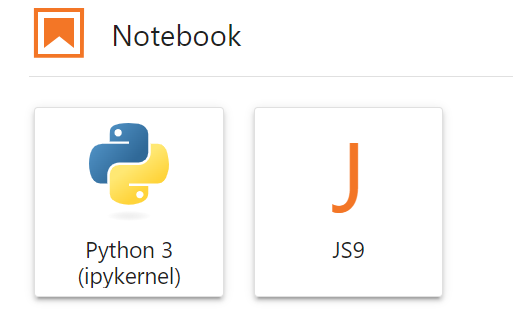}
        \caption{JS9 instance available to be launched separately inside the JupyterLab environment, found under the Notebooks section in the launcher menu.}
        \label{fig:pic3}
    \end{subfigure}
    \vspace{0.5cm} 
    \begin{subfigure}[b]{\hsize}
        \centering
        \includegraphics[width=\hsize]{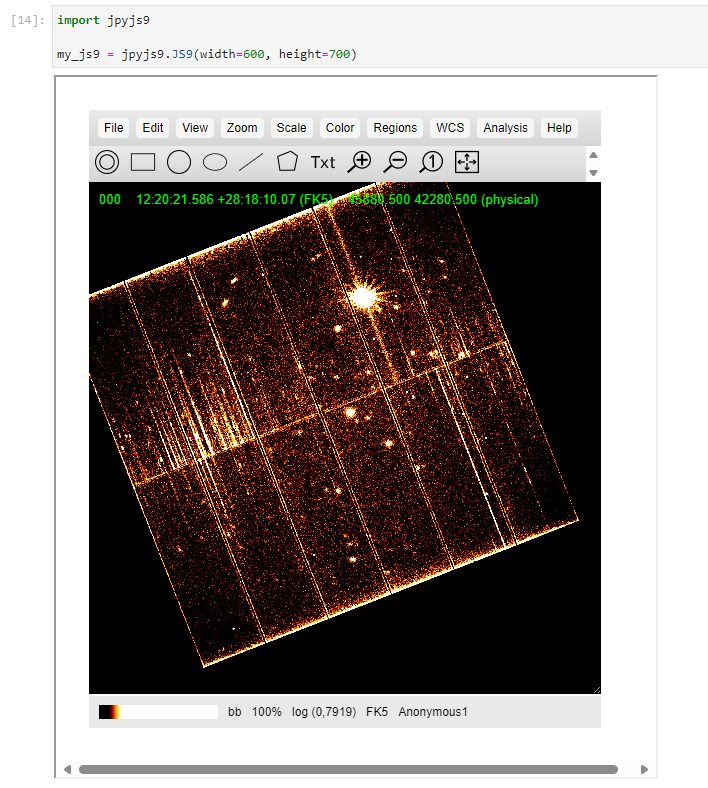}
        \caption{Running JS9 directly inside a Jupyter Notebook importing the library and calling the command on a kernel.}
        \label{fig:pic4}
    \end{subfigure}
    \caption{Two available options for interactive image visualisation inside the XMM-SAS Datalab.}
    \label{js9fig}
\end{figure}

SAS allows for graphical data reduction and analysis. Implementing interactive features into a Jupyter environment presents challenges due to independent cell execution, where parts of the code are run separately, and limited front-end capabilities. To preserve the interactivity of traditional SAS tools, in addition to the standard required Python packages, the user will find tools such as \texttt{Pyjs9} \citep{js9_library}, \texttt{Plotly} \footnote{\url{https://plotly.com}}, and \texttt{lcviz} \footnote{\url{https://github.com/spacetelescope/lcviz}}, all available for interactive visualisation and plotting. These tools enable real-time data manipulation and visualisation, supporting dynamic and iterative analysis processes. The user can find the DS9 application available as a separate Datalab within the provided catalogue, however for continuous analysis flow, we have incorporated the Python \texttt{Pyjs9} package developed by \citep{js9_library} inside the SAS Docker image. Users can access \texttt{JS9} via the launcher area of JupyterLab or import the package to display the application in a Jupyter notebook kernel, as shown in Fig.\ref{js9fig}.

For the interactive lightcurve plotting, the environment does include packages as \texttt{Plotly} and \texttt{lcviz}, which the user can utilize per their needs, however, if required, \texttt{Matplotlib} is also available as an option. We recommend using the package \texttt{lcviz}, as it allows for time range subset selections, where the user can get the values as Python instances for further data analysis when focusing on specific time ranges. The light curve viewer provides multiple tools, such as a combination of multiple light curves, flattening the light curve by removing trends, binning, frequency analysis, and the export of images \citep{lcviz}. The possibility of retrieving the subset ranges allows for a more automated experience when producing spectra, helping preserve the original interactive features that came with SAS (see Fig.\ref{lcvizfig} in Appendix). \texttt{lcviz} is built on top of libraries such a \texttt{Jdaviz} and \texttt{lightkurve}. \texttt{Lightkurve} is used to read the data before plotting, which was built to visualize Tess and Kepler data. To be able to read and plot XMM-Newton light curve data we use a special function, which converts it to a LightCurve() instance, making it readable by \texttt{lcviz} (for more details see section \ref{sec:threads} below and table \ref{tab:pyfunc}).

\subsubsection{SAS analysis threads} \label{sec:threads}

To aid the user, multiple SAS Data Analysis Threads are published and maintained regularly, publicly available in \href{https://www.cosmos.esa.int/web/xmm-newton/sas-threads}{ESA Cosmos} \footnote{\url{https://www.cosmos.esa.int/web/xmm-newton/sas-threads}}. There are five Jupyter Notebook format threads that focus on SAS Start-up and event list manipulation. With the new ESA Datalabs environment, the Jupyter Notebook threads have been transformed as the configuration steps in the new platform are minimized. The available threads are changed to incorporate the packages and tools necessary to include interactivity whilst using pySAS. The files provided as a data volume, are read-only for the user, and if they desire to make changes, each file and folder can be copied to one's permanent workspace. The notebooks available so far include:
 \begin{itemize}
    \item SAS start-up thread in Python,	
    \item How to reprocess ODFs to generate calibrated and concatenated EPIC event lists,
    \item How to filter EPIC event lists for flaring particle background,	
    \item How to extract a light curve and spectrum for an EPIC point-like source,	
    \item How to reduce RGS data and extract spectra of point-like sources.
\end{itemize}
The first four of the threads incorporate all the steps required to do a standard session of SAS analysis from the beginning until the end, replicating the SAS task \texttt{xmmextractor}. 

To make the user experience more streamlined and less code-heavy on the Jupyter Notebook environment, we provide a folder alongside the threads, named \texttt{tools}. This folder includes Python files that have specific predefined functions, which the user can import when running the threads or doing independent analysis, instead of building code from scratch (see the table \ref{tab:pyfunc} for an overview). The user does not have to rely on the provided functions and if desired can explore their own visualisation methods. If interested, the functions can be modified and further developed to match specialised analysis needs. This is where the user can find interactive and non-interactive light curve plotting tools, and options on event file plotting. Furthermore, the interactive tools provided in the SAS datalab allow for the capture and saving of interactive region selections, which subsequently can be used for future scripts.

\section{Science Case Study: Vela X-1} \label{sec:case_study}
To demonstrate the capabilities of the XMM-SAS Datalab, we present a reproducible case study involving the analysis of the X-ray binary Vela X-1. This serves not only as a functionality showcase, but also as a proof of concept for building fully documented, shareable workflows in an open science environment, which can be accessed via git. 

Vela X-1 is one of the best-studied X-ray binary systems, detected in the early years of X-ray astronomy \citep{Chodil_1967b}. Being relatively close in the Galaxy at $\sim$2~kpc distance, and a persistent X-ray source with a bright optical counterpart (HD 77581, visual magnitude $\sim$6.87) it is rather easily observed and shows a rich phenomenology at many wavelengths (see \citet{Kretschmar_2021} for a recent review of the compiled knowledge).

Using the tools and features of the SAS datalab, such as persistent workspaces, git integration, and pySAS, we replicate a published analysis by \citep{Diez_2023} starting from data retrieval and processing to final spectrum production. Furthermore, we compare both the spectra from the original research and the ESA Datalabs replica to test the reliability of the environment.
This has been done in a Jupyter notebook format serving as a comprehensive step-by-step guide demonstrating retrieval and processing of XMM-Newton data for the chosen source. Access to this material can be found at the public git repository \href{https://github.com/esa/xmm-sas-datalabs-case-study}{XMM-Newton SAS in ESA Datalabs} \footnote{\url{https://github.com/esa/xmm-sas-datalabs-case-study}}. The JupyterLab structure of ESA Datalabs, with its git integration, notebooks, and persistent workspaces, allows users to package complete reproducible workflows. This makes it easier to publish and share fully documented analyses via repositories as the one mentioned above. 

In this example, we do not go into the specifics of spectral fitting, the reader is advised to check out the designated paper for the scientific details. The purpose of the paper is not to discuss any science. All results are shown only for the purpose of presenting SAS flow inside ESA Datalabs. 

\subsection{Method \& Set-up}
To match the original analysis, we chose observation ID 0841890201, which occurred between the 3rd and 5th of May, 2019. For event file generation, SAS version 21.0 is used alongside with CCFs from September 2024 onwards. The Observation Data Files (ODFs) are accessed and downloaded using the task \texttt{startsas}. During the observation period, the XMM-Newton instruments used included EPIC-pn, EPIC-MOS, and RGS. The total exposures are five with 3 EPIC and 2 RGS. During the observation, the intrinsic luminosity of the source was approx $1.5 \times 10^{36} \, \rm{erg} \, \rm{s^{-1}}$ in the 0.5--10 keV energy band \citep{Diez_2023}. This is a very bright source; hence EPIC-pn exposure was set in timing mode. In this mode, the arrival times of individual X-ray photons are recorded with high time resolution. Rather than focusing on spatial information, there is more emphasis on the temporal aspects of X-ray events, making it ideal for studying time-dependent phenomena. Hence we run \texttt{epproc} for the timing mode to get the observation event files. According to \cite{Diez_2023}, the default calibration is not used for the timing mode, which otherwise could cause a shift to higher energies in the instrumental and physical source lines. Hence, the Rate-dependent PHA is excluded, whilst Rate-dependent CTI correction is applied. The command \texttt{epproc} is executed with the following arguments:

\begin{lstlisting}
command  = 'epproc'
inargs   = 'withdefaultcal=N', 'withrdpha=N', 'runepreject=Y', 'withxrlcorrection=Y', 'runepfast=Y'
\end{lstlisting}

No strong flaring particle background was found and no X-ray background was extracted due to the very high luminosity of Vela X-1 illuminating the whole focal plane. A barycentric correction is applied using the \texttt{barycen} task to correct the light curve times to the solar system barycenter reference frame. The source region is selected as \texttt{RAWX in 32:44} in pixels for the square rectangle region. The event filter selects the energy range as \texttt{PI in 500:10000} in eV and includes single and double pixel pattern events with \texttt{PATTERN<=4}. The selected source region and event filtering are taken from \cite{Diez_2023}, specifically chosen to avoid possible background noise or scattering effects from dust. The task \texttt{evselect} is run with the above-mentioned filtering conditions with a time bin size of 283 sec. The bin value is selected to match the pulse period of $P=283.4447 \pm 0.0004$ s derived in \cite{2022A&A...660A..19D}. We applied the task \texttt{epiclccorr} to perform absolute (events lost through detector inefficiencies) and relative (varying with time) corrections on the lightcurves.

Each incoming X-ray photon is converted into an electronic signal. If multiple photons arrive at the same detector location (pixel or nearby pixels) during the integration time rapidly, their signals overlap, leading to an inaccurate representation of the detected X-ray events, which is called pile-up. This can distort and impact the accuracy of the measured X-ray spectra and light curve, leading to a false output of higher flux at high energies. The SAS command \texttt{epatplot} plots EPIC-pn event pattern statistics, which helps the user to see if pile-up has occurred. Therefore, the source region is selected from which the piled-up regions have been removed (see Fig.\ref{fig:epatplot} in \ref{app:epatplot}). Hence, the spectrum is produced for the new source region with the core of the PSF coordinates as 36 and 40 excised between the 37 and 39 column coordinates (\texttt{RAWX in [{32}:{36}] || RAWX in [{40}:{44}]}).

\begin{figure}
    \centering
    \includegraphics[width=\linewidth]{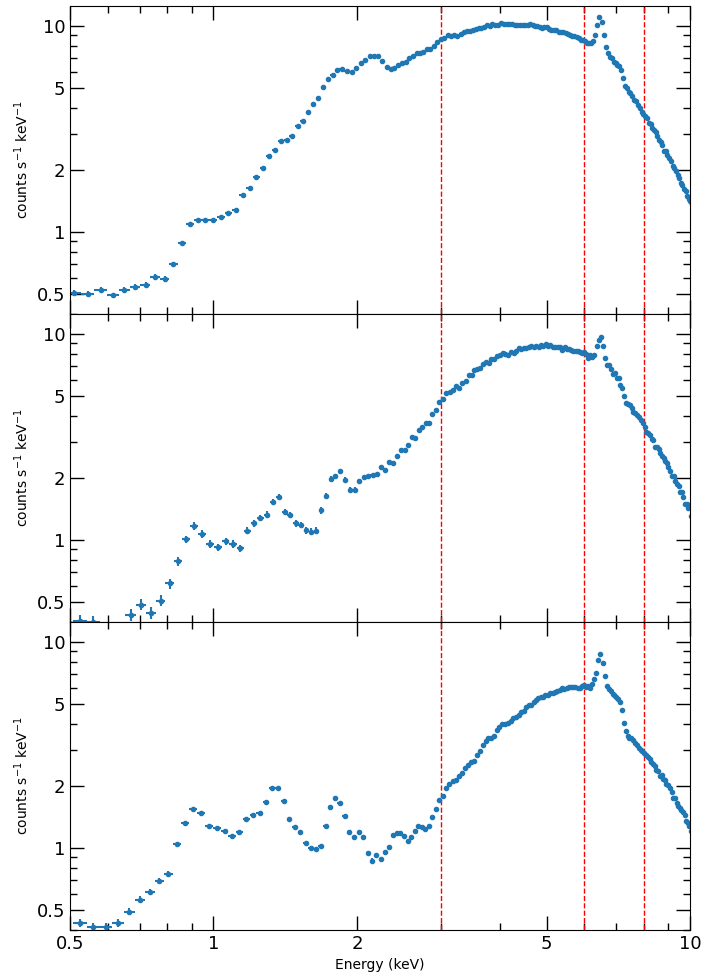}
    \caption{Spectra extracted for the three distinct phases of observation. They are presented in a chronological order from top to bottom. The vertical dashed lines are the borders of the four energy bands chosen by \cite{Diez_2023} for extraction of energy resolved light curves.}
    \label{fig:3phasesepctra}  
\end{figure}

\subsection{Results}

We visualise the 0.5-10 keV XMM-Newton EPIC-pn light curve to observe the general behaviour of Vela X-1, as it shows high variability in flux. The major flares observed (see Fig.\ref{fig:lightcurvetimeresolution283} in \ref{app:lightcurve}) match those presented in \cite{Diez_2023} Fig. 2. A characteristic long-lasting flare can be detected roughly from $T \approx 0.4$ d, which lasts for approximately $8$ ks. The observation is split into three distinct phases. The first one is from the start of observation until $T \approx 0.57\mathrm{d}$ (i.e. Orbital Phase $\lesssim 0.46$) with a stable hardness ratio, the second phase lasts until $T \approx 0.76\mathrm{d}$ (i.e. $0.44 \lesssim $ Orbital Phase $ \lesssim 0.46$), and the final phase is characterised with a rise of the hardness ratio until the end of the observation (i.e. $0.46 \lesssim $ Orbital Phase). We select the observation time ranges, characterised by the hardness ratio change as stated in \cite{2022A&A...660A..19D}, where the phenomenon can be observed in the $3.0-5.0$ keV energy band.

Using the phases mentioned above, three different EPIC-pn spectra are extracted and presented in Fig. \ref{fig:3phasesepctra} with chronological order. Looking at the spectrum on the last panel, a significant number of emission lines are visible, which is revealed as the continuum is suppressed by the absorption from the stellar wind \citep{Watanabe_2006, Lomaeva_2020}. To investigate the variability of Vela X-1 it is possible to extract pulse-by-pulse spectra throughout the observation. Instead of manually entering selected time ranges, we define a loop to select Good Time Intervals (GTI) for each pulse increase of $P=283.44$ s from the beginning until the end of observation time. For each GTI file, the script produces the corresponding spectrum, resulting in 392 spectra. For further details on the scripts please refer to table \ref{tab:pyfuncvela} in \ref{app:scriptsforvela}.

\begin{figure*}
    \centering
    \includegraphics[width=\linewidth]{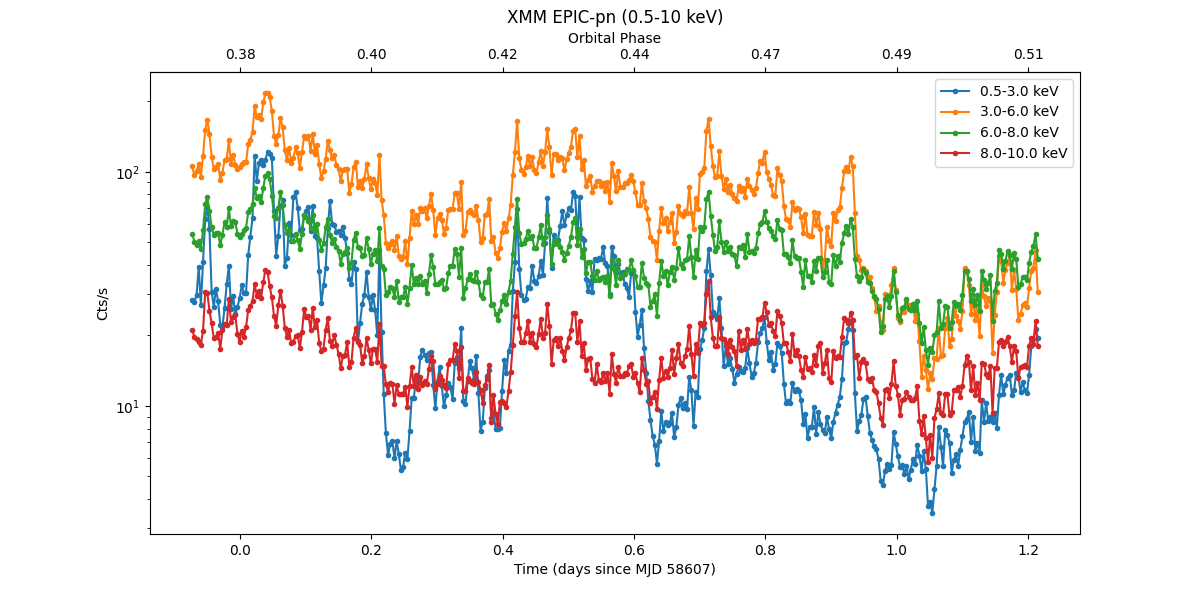}
    \caption{Light curves extracted for four different energy bands of $0.5-3.0$ keV, $3.0-6.0$ keV, $6.0-8.0$ keV, $8.0-10.0$ keV. The light curves have been extracted after pile up correction. This is a match of Fig. 4 from \cite{Diez_2023}.}
    \label{fig:timeresolvedLC}  
\end{figure*}

\cite{Diez_2023} chose four distinct energy bands of interest for further investigation, from analysis of the last panel in Fig.\ref{fig:3phasesepctra}. This is done to compare the count rate in different energy bands to check for any variability. The first energy band is chosen to range from 0.5 keV to 3.0 keV, to cover the low-energy emission lines, next range is between $3.0-6.0$ keV, further observing the low-energy sections of the continuum without the emission lines. The third energy range focuses on the prominent Fe K$\alpha$ emission line at $6.0-8.0$ keV. The final band contains the high energy section at $8.0-10.0$ keV. Light curves are extracted from the corresponding energy bands and displayed in Fig. \ref{fig:timeresolvedLC}. Each band exhibits similar features and variability. The difference arises at higher flaring activity in lower energy bands, where there is a bigger gap reaching comparatively lower count rate.

\begin{figure*}

    \begin{subfigure}[t]{.49\textwidth}
        \centering
        \includegraphics[width=\linewidth]{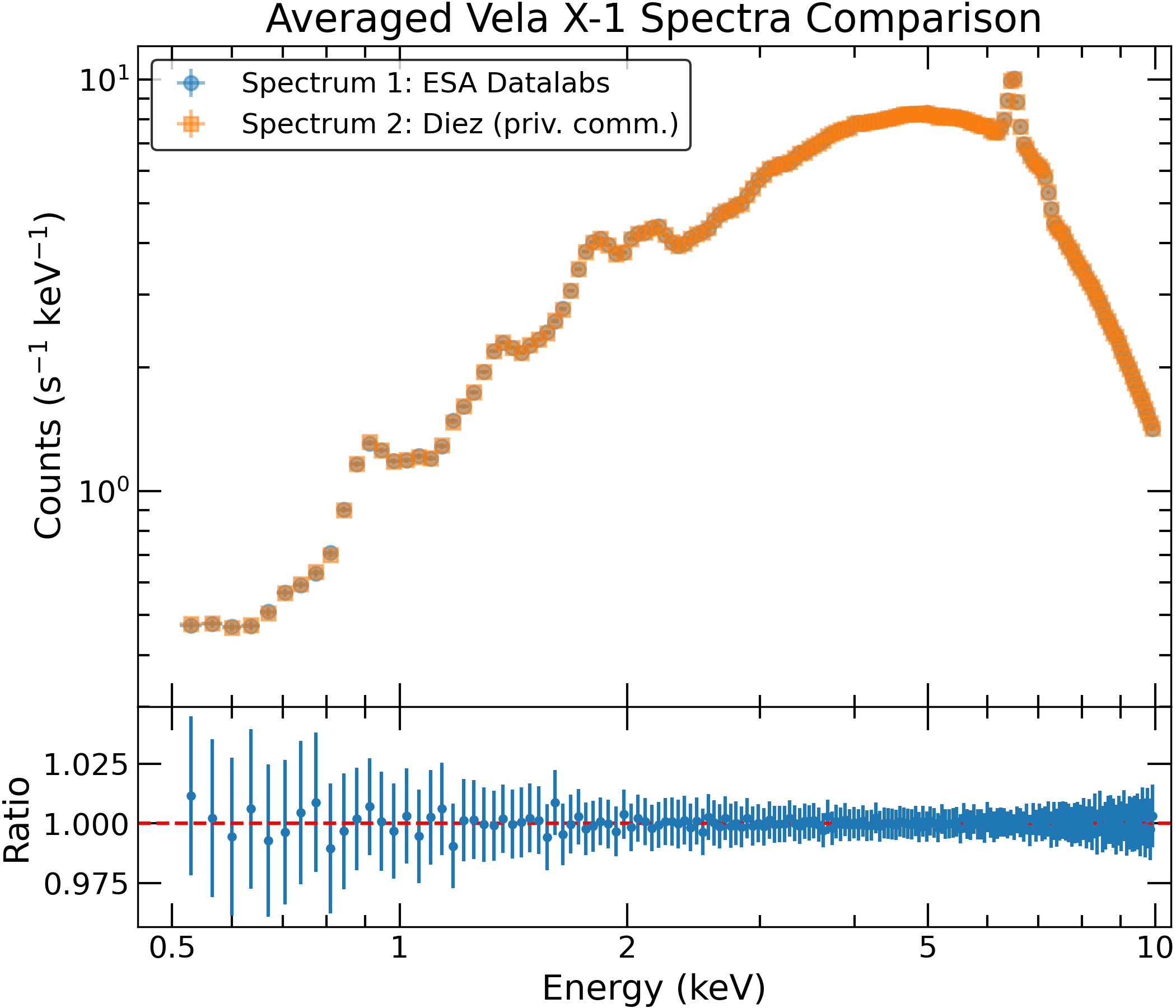}
        \caption{The top panel shows overlap of both averaged spectra of Vela X-1. The bottom panel displays the ratio difference between the two spectra.}
        \label{fig:spectra_comparison}
    \end{subfigure}
    \hfill
    \begin{subfigure}[t]{.49\textwidth}
        \centering
        \includegraphics[width=\linewidth]{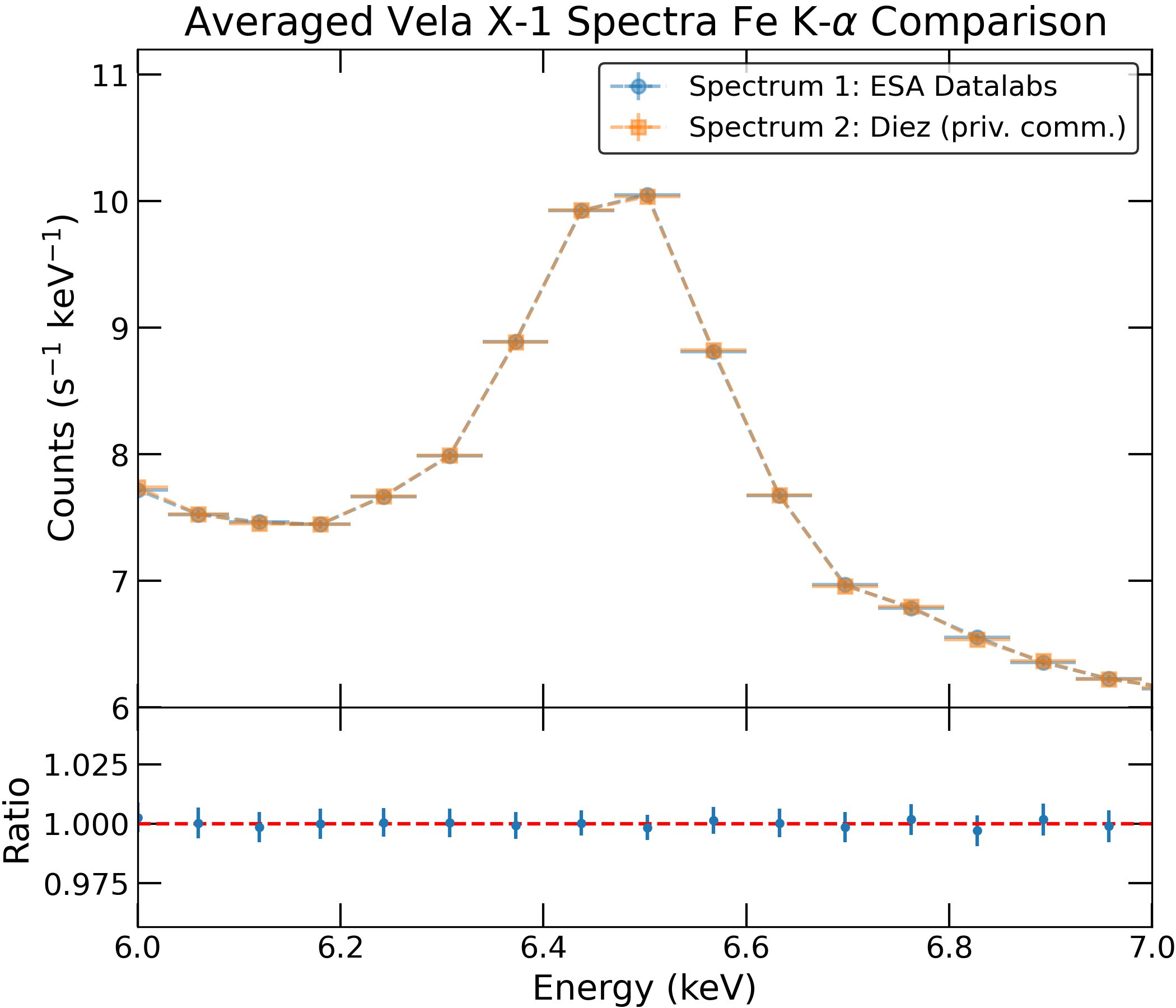}
        \caption{The upper graph highlights the Fe K$\alpha$ emission line overlap between the two spectra, both of which exhibit the same emission shape. The lower graph presents the ratio between the spectra for the selected iron emission line.}
        \label{fig:Fe_line_comparison}
    \end{subfigure}
    
    \caption{Plots of both averaged XMM-Newton EPIC-pn spectrum of Vela X-1, extracted on the Datalabs platform and spectrum produced by C. M. Diez (priv. comm., 2024) extracted through local SAS set-up used for the paper \citep{Diez_2023}.}
    \label{fig:averagedspec}
\end{figure*}

To get a general view, a time-averaged spectrum is visualised with pyXspec and Matplotlib, see Fig. \ref{fig:averagedspec}. Looking at the spectra presented in Fig.~\ref{fig:spectra_comparison} we can reproduce the same time-averaged spectrum in agreement with \cite{Diez_2023}'s findings using their data extraction method within the Datalabs environment. The time-averaged spectrum of the original research is obtained through private communication with C. M. Diez. Comparison of both spectra shows an overlap, even within a closer look at the characteristic iron (Fe K$\alpha$) line feature (Fig.\ref{fig:Fe_line_comparison}). To quantify the deviation, a residual graph is plotted, showing the ratio between the original spectrum and Datalabs spectrum, whilst applying interpolation. The mean from the no deviation line of 1 is calculated to be $\overline{\delta} = 0.0018$ for data points between 0.5 and 10.0 keV. These results demonstrate that the ESA Datalabs platform accurately reproduces the original local SAS output, with only minimal deviations in the spectra. 

Though deviations may occur due to numerical precision or rounding methods used on different platforms, the primary source of discrepancy is likely due to differences in calibration files. Specifically, the original study used SAS version 20.0 with April 2022 CCFs, while our replication uses SAS 21.0 and CCFs from October 2024.
Between April 2022 and October 2024, several updates were made to the XMM-Newton Calibration Documentation, particularly affecting the EPIC instruments. These changes could influence spectral results. In June 2022, the EPIC Calibration Status was updated to Version 3.13, providing the latest calibration information up to that date. Subsequently, in April 2024, the EPIC Calibration Status was updated to Version 3.14, incorporating calibration data up to October 2024, reflecting the most recent adjustments and improvements. Key changes in the CCFs during this period include updates to Response Matrix Files (RMFs), which can affect spectral fitting, especially in the soft X-ray band. Differences in RMFs between April 2022 and October 2024 may lead to variations in spectral modeling and parameter estimation. Modifications to Effective Area Files influence the instrument's sensitivity across different energy bands, and changes in these files can alter the observed fluxes and spectral shapes. Regular monitoring and updates to time-dependent calibration parameters, such as detector gain and background models, ensure accurate spectral analysis. Differences in these parameters between the two periods could lead to discrepancies in the spectral results. Additionally, the release of XMM-CCF-REL-407 on March 15, 2024, updated the long-term Charge Transfer Inefficiency (CTI) correction for EPIC-pn, which could affect spectral fitting and parameter estimation \citep{XMM-Newton_CCF_Release_Notes}.

The results and scripts of the case study highlight how XMM-SAS Datalab can serve as a foundation for creating reproducible research packages. The complete pipeline, from data download to analysis, is captured in a single Jupyter notebook and supporting scripts, all available in a public git repository \footnote{\url{https://github.com/esa/xmm-sas-datalabs-case-study}}. By combining cloud-based tools, interactive scripts and notebooks, direct access to up-to-date CCFs, and the latest available SAS version, the datalab allows for transparent and shareable workflows. Although SAS datalab does not ensure that the same SAS software versions and CCFs would be available for full reproducibility of research scripts, strict application of the same analysis flow is ensured, while software and calibration will be kept up-to-date, as in the example of the Vela X-1 case study. This approach aligns with growing demands for reproducibility in astrophysics and offers a model that can be extended to future X-ray missions or multi-wavelength collaborations.

\section{Discussion} \label{sec:discussion}
 
XMM-Newton has seen continuous advancements in technology and user needs over time. As a result, new opportunities to improve the user experience have emerged. SAS is an example of such a tool that has evolved significantly over the years to meet these demands. Recently, SAS has expanded to include a cloud-based option, accessible via a web browser, which allows teams to collaborate efficiently by sharing their work in a central location. This evolution reflects the ongoing adaptation of the software to better support the scientific community. The shift from local desktop analysis to the cloud addresses long-standing challenges related to the accessibility of scientific tools, particularly by enabling remote access to resources. This shift enhances collaborative research in X-ray astronomy by allowing teams to work in a shared environment, regardless of their physical location, and facilitating the seamless exchange of data and results in real time.

\subsection{Case Study Results \& Implications}
The Vela X-1 case study demonstrates that complex and diverse SAS analysis tasks can be executed within ESA Datalabs using pySAS at a level of precision and reliability comparable to a local SAS setup. The aim of replicating the original research was to validate the platform's capabilities by converting traditional shell-based SAS workflows into Python scripts. The recreation of \cite{Diez_2023} demonstrates the platform's capabilities.

By reproducing each step of the original analysis, including data retrieval, calibration, event file filtering, and spectrum extraction, the study highlights the integration of pySAS into the Datalabs environment. Additionally, the scripts provide new users with a comprehensive starting point for XMM-Newton data analysis. These resources can be directly imported and executed in the JupyterLab interface of the XMM-SAS Datalab, allowing for straightforward replication of the case study. A lack of deviations with discernible systematic pattern when comparing Datalabs and local SAS results, further reinforce the platform’s reliability for scientific research. Therefore, the direct translation of traditional SAS tasks into Python scripts not only confirms the flexibility and accessibility of the new system but also emphasizes its potential to enhance reproducibility and integration with other analysis tools.

\subsection{Benefits vs. Trade-offs} 
XMM-SAS Datalab not only integrates SAS with a cloud platform, but also provides interactive notebooks, tutorials, and Python tools tailored for XMM data analysis. Its direct connection to ESA archives ensures direct access to up-to-date calibration files.
Compared to the previous options, in the Datalabs environment SAS can be used collaboratively, providing online space to store files and scripts. This allows users to access their work from any device with internet connectivity. 
The collaborative features of the Datalab, such as shared workspaces and integrated data volumes, offer possibilities for team-based research and multi-wavelength studies. Researchers can build and share their custom environment and scripts on top of the provided SAS image, thereby creating complex analysis flows and pipelines. Furthermore, the availability of pre-configured analysis tools and examples supports both experienced researchers and newcomers to X-ray astronomy, reinforcing the platform’s role as a valuable resource for the scientific community.

However, there are trade-offs in switching to ESA Datalabs. Working offline is not possible, hence you cannot access your stored data or do any SAS analysis, unlike a local SAS installation. This also means that access to the platform, your data, and SAS, is affected by service updates or maintenance. This makes the SAS user dependent on external factors and internet connection. Furthermore, using XMM-SAS Datalab requires Python and pySAS knowledge to access interactive SAS features. The Docker interface is that of a JupyterLab, which does provide for the usage of a terminal, but lacks the previous GUI involved at the local terminal, however this does not prevent researchers from running SAS in a traditional format. 
Overall, ESA Datalabs provides benefits, that surpass the potential challenges. It also should be noted that this is a platform at its early stages, and has much room for development as more people use it, just as SAS has over the years.

\subsection{Comparison with other services}
Besides ESA Datalabs other e-science platforms can be used for collaborative and accessible research. One of those platforms that also includes SAS is SciServer \citep{Taghizadeh_Popp_2020}. Similar to ESA Datalabs, SciServer users have access to a private storage area and can create and run their own Docker container selecting data volumes of interest and a compute image. SciServer also offers an integrated environment that allows users to analyze large datasets directly in the cloud, removing the need for local hardware. Furthermore, much like ESA Datalabs, it provides a collaborative workspace where users can share data, tools, and workflows, enhancing team-based research. SciServer is open to the public. 

To run SAS in SciServer, the user can create a new Docker container, selecting a compute image that includes XMM-SAS, such as HEASARC. The user can then upload or access their data from the private storage area or shared data volumes, and run XMM-SAS directly within the container with JupyterLab interface for data reduction and analysis. The platform's integrated environment allows the user to perform all these tasks without needing to install any software locally. This gives the user an alternative platform to process XMM-Newton data. Though built on similar ideas with both providing access to SAS in a JupyterLab environment, both platforms exhibit some differences. Whilst SciServer might be preferable, especially for usage of other X-ray data from different missions, ESA Datalabs provides a service that is specifically focused on XMM-Newton usage only, with tutorial notebooks and Python tools, that are exclusive to the SAS datalab. Currently, XMM-SAS Datalab is not configured for other missions such as XRISM or Chandra. However, users can independently create new environments for other data types using the Datalabs Editor. We acknowledge the benefits of each platform and encourage the user to try both.

\section{Summary \& Conclusions} \label{sec:conclusion}
The XMM-SAS Datalab presents a step in adapting X-ray data analysis to modern cloud-based research environments. By eliminating the complexities of SAS setup and integrating it into the ESA Datalabs platform, researchers can now access a Jupyter-based interface that supports collaborative workflows and facilitates streamlined data processing. 

Looking ahead, the SAS datalab's potential lies in its adaptability and future updates. As the platform gains wider use, systematic feedback from the user community is expected to drive further enhancements for ESA Datalabs and SAS. Such improvements will aim to refine data processing accuracy and expand the platform’s capabilities for both educational and advanced research purposes. This initiative illustrates how legacy software can be integrated into modern e-science platforms, paving the way for broader applications in multi-mission and multi-wavelength astronomy.

\section*{Acknowledgements}
We thank everyone who contributed to the development and testing of the XMM-SAS Datalab. Special acknowledgements to European Space Astronomy Center, ESA Data Science team, XMM-Newton SOC, and ESA Datalabs team. Further thanks to XMM-Newton GOF, R. Tanner, D. Nguyen, and A. Zoghbi.

\appendix
\onecolumn
\section{SAS Setup \& Requirements} \label{setup appendix}
\subsection{Downloadable Products}

\begin{table}[ht]
\centering
\caption{SAS 21.0 Downloadable Products. Table from \href{https://www.cosmos.esa.int/web/xmm-newton/sas-download}{XMM-Newton SAS Download page}: \url{https://www.cosmos.esa.int/web/xmm-newton/sas-download}.}
\begin{tabular}{|l|l|l|l|}
\hline
\textbf{Build/Product} & \textbf{Kernel Version} & \textbf{libc Version} \\ \hline
Linux Ubuntu22.04LTS & 5.15.0 & 2.35  \\ \hline
Linux Red Hat 8.6 & 4.18.0 & 2.28  \\ \hline
Linux Ubuntu20.04LTS & 5.15.0 & 2.31  \\ \hline
Linux Ubuntu18.04LTS & 4.15.0 & 2.27  \\ \hline
Linux Centos7.3 & 3.10.0 & 2.17  \\ \hline
macOS 12.6 or Monterey (Darwin 21.6.0) & Darwin 21.6.0 & 1311.120.1  \\ \hline
VM4SAS21.0 (Linux Ubuntu 22.04LTS Virtual Machine for VMWare) & 5.15.0 & 2.35  \\ \hline
docker4sas:sas\_21\_0 (SAS Docker Image based on Linux Ubuntu22.04LTS) & 5.15.0 & 2.35  \\ \hline
\end{tabular}
\label{tab:sas_products}
\end{table}

\subsection{Software Requirements}

\begin{table}[ht]
\centering
\caption{Software and Tools Required to Work with SAS 21.0. Table from \href{https://www.cosmos.esa.int/web/xmm-newton/sas-requirements}{XMM-Newton SAS Requirements page}: \url{https://www.cosmos.esa.int/web/xmm-newton/sas-requirements}.}
\begin{tabular}{|l|l|p{11cm}|}
\hline
\textbf{Name} & \textbf{Version}  & \textbf{Comments} \\ \hline
X11 & Provided by building OS & On macOS, you need to install XQuartz regardless of the macOS version.  \\ \hline
Python & 3.10.6  & The following Python packages are required to work with SAS 21.0: astropy, numpy, matplotlib, requests, pyds9, PyQt5, beautifultable, scipy, pypdf, notebook, astroquery>=0.4.3 (v0.4.3 is mandatory !). \\ \hline
Perl & 5.34.1  & The following Perl modules are required: Switch.pm, Shell.pm, CGI.pm. \\ \hline
ds9 & 8.4.1 or later & The tool xpa 2.1.20 (or later), not included with ds9, is required to work with SAS. Please download it from the same site and install it together with ds9. For macOS, there are two ds9 versions: one based on X11 and another for Aqua Windowing. Use the X11 version. \\ \hline
Grace & 5.1.25  & Installation from source code is difficult. CentOS and Ubuntu provide binary packages that are easy to install. For macOS, install Qt Grace 0.2.6.  \\ \hline
Heasoft & 6.30  &  To build Heasoft from source code and make such build fully compatible with SAS 21.0, please do not forget to define both, the PERL and the PYTHON environment variables, to point to the Perl 5.34.1 and the Python 3.10.6 (or later) binaries used to work with SAS 21.0. \\ \hline
WCSTools & 3.9.7  & Easy to build from source. Do not install WCStools v3.9.5 because its \textit{scat} tool does not work when downloading data, e.g., by setting the UB1\_PATH environment variable to query the USNO Catalogue. \\ \hline
\end{tabular}
\label{tab:sas_software}
\end{table}

\newpage

\section{SAS Analysis Threads}
\begin{figure}[h]
    \begin{subfigure}[a]{\hsize}
        \centering
        \includegraphics[width=18 cm]{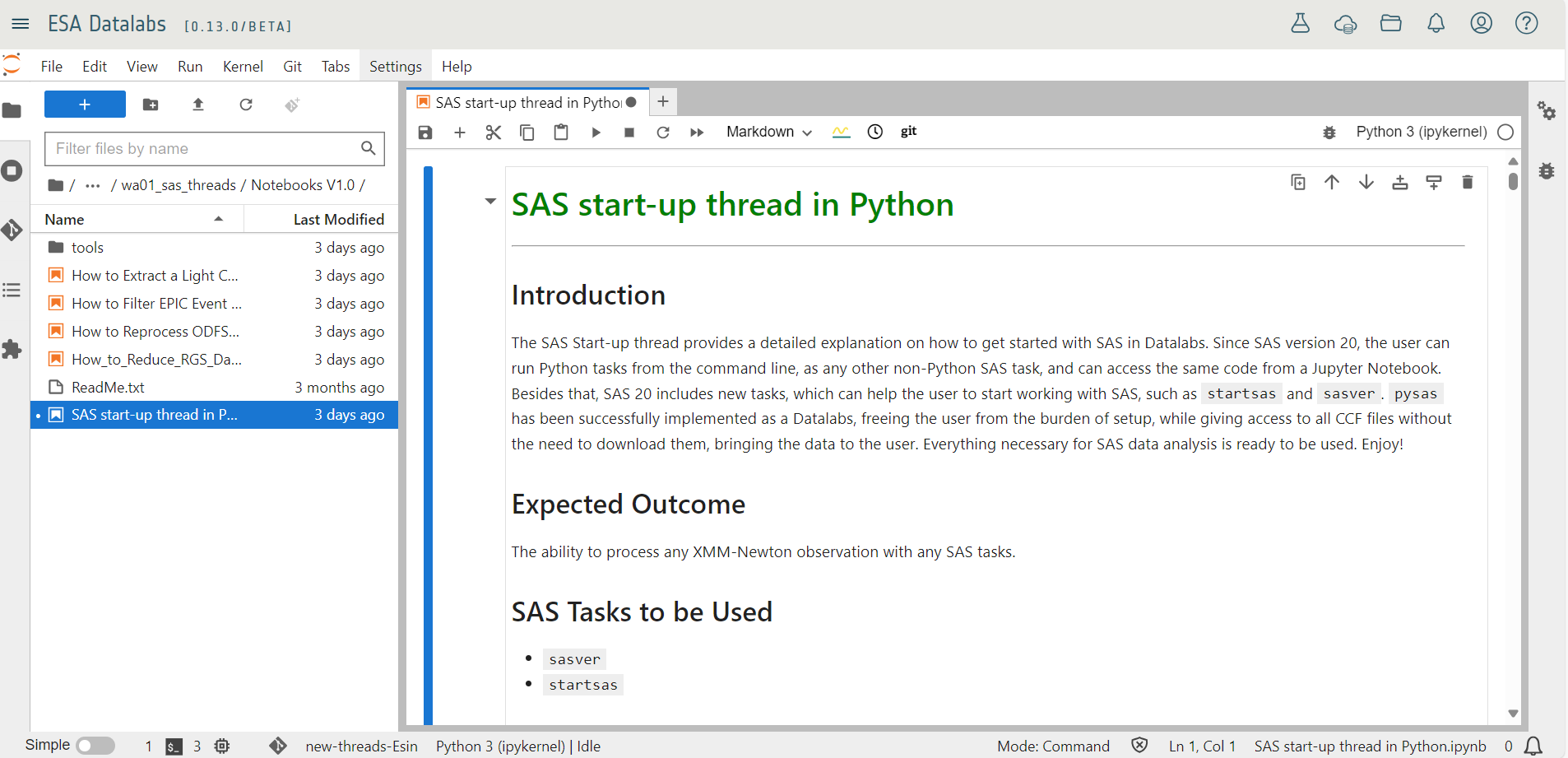}
    \end{subfigure}
    \vspace{0.5cm}
    \begin{subfigure}[a]{\hsize}
        \centering
        \includegraphics[width=18 cm]{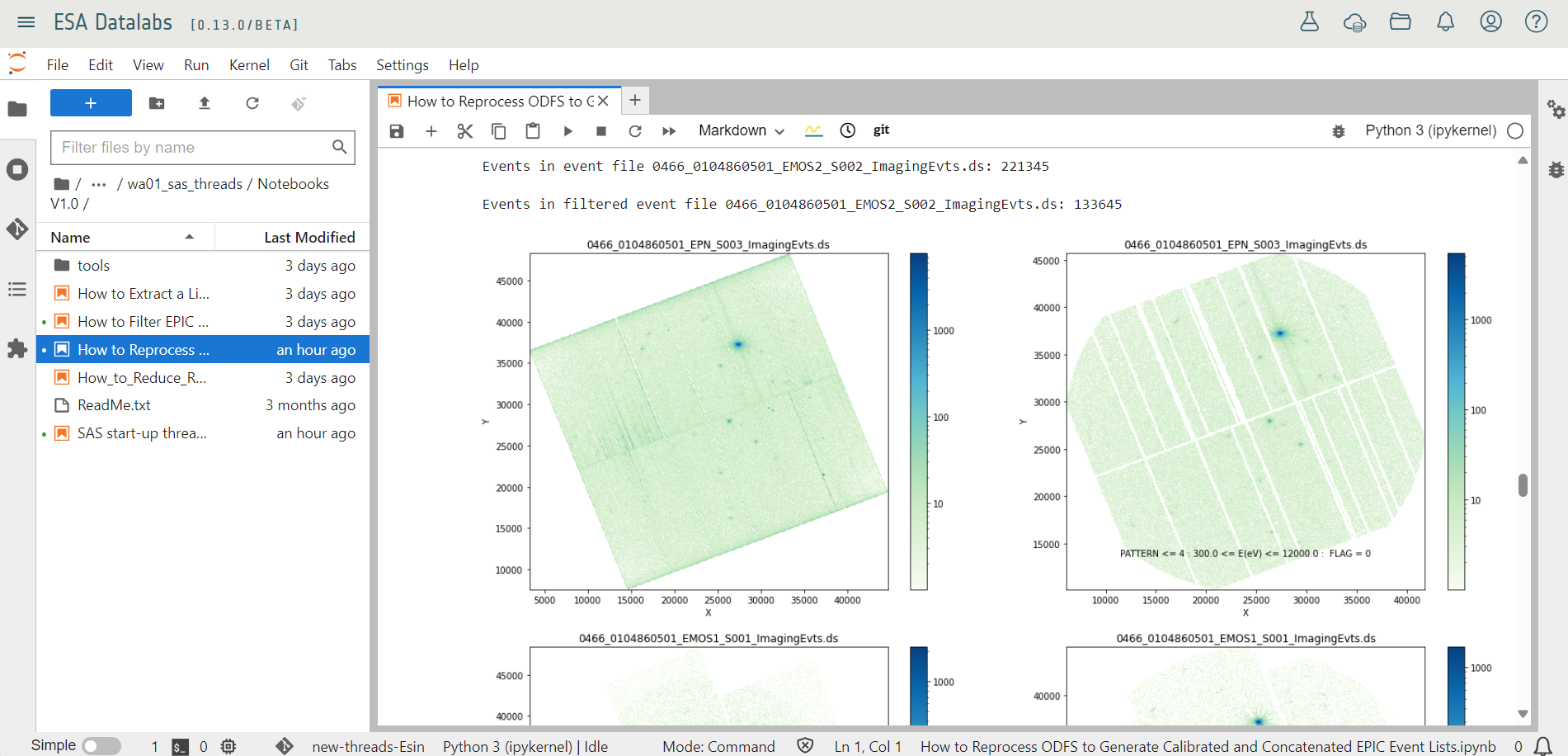}
    \end{subfigure}

    \caption{Images displaying sections of the available SAS analysis threads in Datalabs, which can be run through the available JupyterLab environment.}
    \label{fig:threads}
\end{figure}

 Five SAS analysis threads have been made available in ESA Datalabs as a data volume content. Alongside them a tools folder is provided with useful Python functions that help with interactive features and visualisation, whilst allowing for a tidier Jupyter notebook work page. An example outlook into some of the threads can be seen in Fig. \ref{fig:threads}. For the content description of the tools folder please refer to table \ref{tab:pyfunc}.

\newpage

\section{Available functions} 
\begin{figure}
   \centering
   \includegraphics[width=\linewidth]{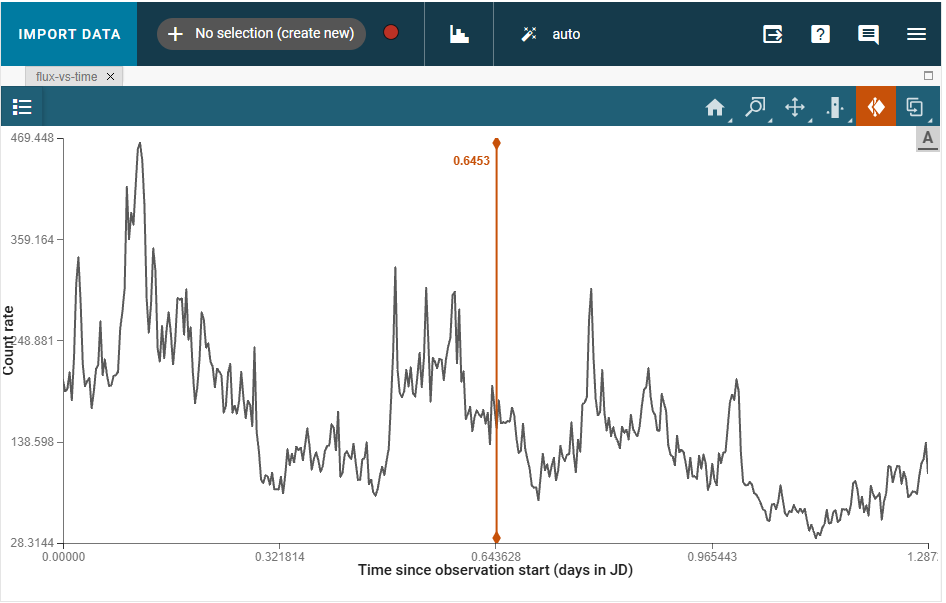}
      \caption{A kernel within the XMM-SAS Datalab visualising the corrected lightcurve from Fig. \ref{fig:lightcurvetimeresolution283}. A marker blue line can be moved interactively, which shows exact coordinates. The red marked subset is an interactively chosen time range subset, where the user can get the values as a Python instance.}
         \label{lcvizfig}
   \end{figure}

SAS analysis threads are provided in ESA Datalabs, alongside a tools folder, containing custom made Python functions. Each Python file holds designated functions for different purposes, where the focus mainly is on interactive visualisation. The available functions are listed in table \ref{tab:pyfunc}. An example output of the functions \texttt{read\_lightcurve} and \texttt{lcviz} can be seen in Fig. \ref{lcvizfig}.

\begin{table}[h]
    \centering
    \begin{tabular}{|c|c|p{8cm}|}
        \hline
        \textit{Python file} & \textit{Functions} & \textit{Description} \\
        \hline
        \texttt{plot\_events.py} & 1. \texttt{plot\_events} & 1. Reads an event file, applies specified filters, and generates plots to visualize the events based on given criteria, optionally displaying both raw and filtered data. \\
         & 2. \texttt{gti\_filtered\_image} & 2. Reads an event file and its corresponding Good Time Interval (GTI) filtered file, and generates side-by-side plots to visualize the events, allowing for comparison between the original and GTI-filtered data. \\
        \hline
        \texttt{plotLC.py} & 1. \texttt{plotLCmatplotlib} & 1. Processes multiple FITS files to extract and plot light curves using Matplotlib, optionally applying a threshold and coordinate limits, and annotates the plots accordingly.\\
         & 2. \texttt{plotLC} & 2. Uses Plotly to generate interactive light curve plots from multiple FITS files, applying thresholds if provided, and labels the plots with file-specific names.\\
          & 3. \texttt{read\_lightcurve} & Reads a light curve FITS file, processes the data to ensure correct units and column names, and returns a LightCurve object compatible with the Lightkurve library, enabling it to be loaded and visualized in the LCviz tool for interactive light curve analysis.\\
          & 4. \texttt{lcviz} & Visualizes multiple light curves using the LCviz viewer, allowing users to load and display light curve data with optional labels. It ensures that the number of labels matches the number of light curves, and can also add a synthetic light curve with a constant flux threshold if specified. The function customizes the axis labels and plot appearance before displaying the visualization.\\
        \hline
        \texttt{plotRGS.py} & \texttt{plotRGS} & Generates two subplots from a FITS file: the first shows a 2D histogram of photon energy versus wavelength with overlaid source region data, and the second displays a convolved 2D histogram of cross-dispersion versus wavelength, including spatial and background region overlays for a specified source.\\
        \hline
        \texttt{js9helper.py} & 1.\texttt{visualise} & Displays a FITS image in the JS9 interface and allows users to modify the scale, colormap, contrast, and bias for enhanced visualization. \\
         & 2.\texttt{getRegions} & Retrieves and processes region data from the JS9 interface, identifying source and background regions and extracting their coordinates and sizes in various coordinate systems, while saving information in a file.\\
        \hline
    \end{tabular}
    \caption{Description of Python functions made available to the user for a swifter pySAS analysis, eliminating the need for the user to come up with their own visualisation and plotting functions.}
    \label{tab:pyfunc}
\end{table}

\newpage

\section{Science Case Study: Vela X-1}

\subsection{Pile-Up Check}\label{app:epatplot}
The SAS command \texttt{epatplot} can be used to visualise whether a bright source is affected by pile-up effects. Fig.\ref{fig:epatplot}, presented below, shows that the source region selections for Vela X-1 data exhibit pile-up effects, as there are more double events and fewer single events than predicted. Effects of pile-up are discarded by re-adjusting source regions and excluding the brightest centre coordinates. 
\begin{figure}[H]
    \begin{subfigure}[t]{.49\textwidth}
    \centering
    \includegraphics[width=\linewidth]{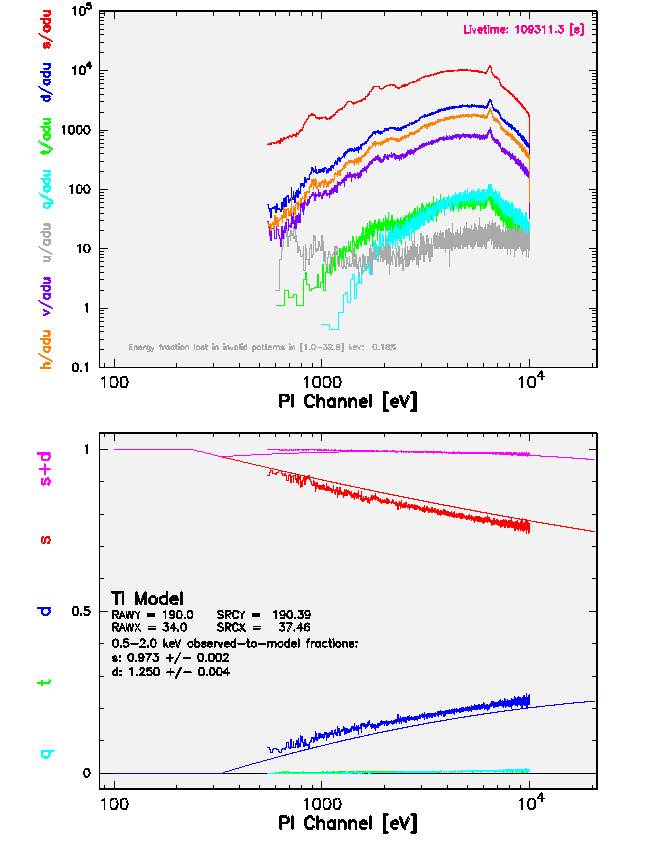}
  \end{subfigure}
  \hfill
  \begin{subfigure}[t]{.49\textwidth}
    \centering
    \includegraphics[width=\linewidth]{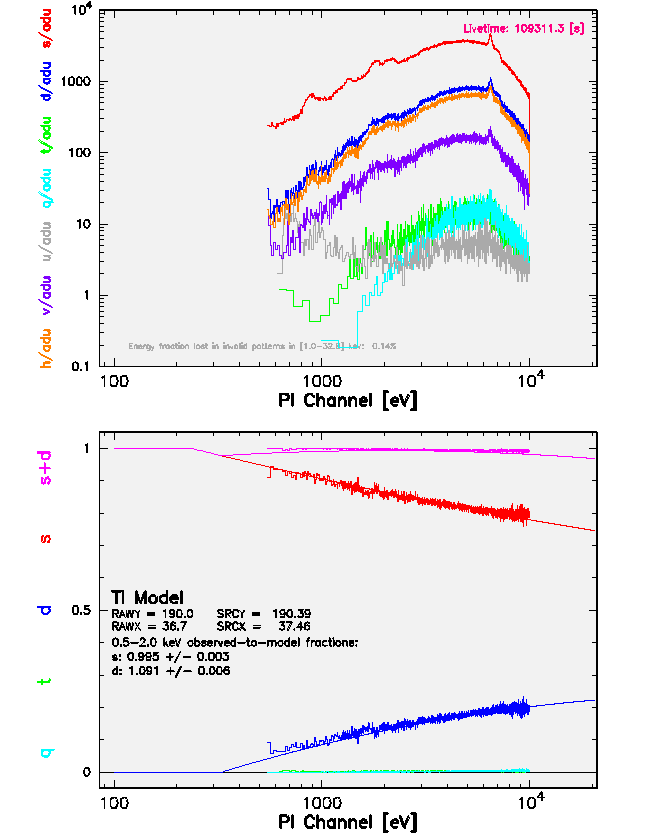}
  \end{subfigure}
  \caption{Output of SAS task \texttt{epatplot} in XMM-SAS Datalab. The blue line represents double events, while the red one is for single events. The smooth lines displayed in the bottom graphs show predicted theory lines. The graph on the left panel shows the effects of pile-up on the detectors, with a higher number of double events than predicted. The right panel displays the task output after the selection, where the theoretical line aligns with the observation, mitigating pile-up effects, please beware of the y-axis change.}
  \label{fig:epatplot}
\end{figure}

\newpage

\subsection{Visualising the Light Curve}\label{app:lightcurve}
After source and background region selections, we can extract the average light curve to study the variable behaviour and flares of Vela X-1 visualised at the energy range of $0.5-10$ keV in the Fig.\ref{fig:lightcurvetimeresolution283} below. SAS command \texttt{epiclccorr} is used to perform correction of the source light curve.

\begin{figure}[H]
    \centering
    \includegraphics[width=\linewidth]{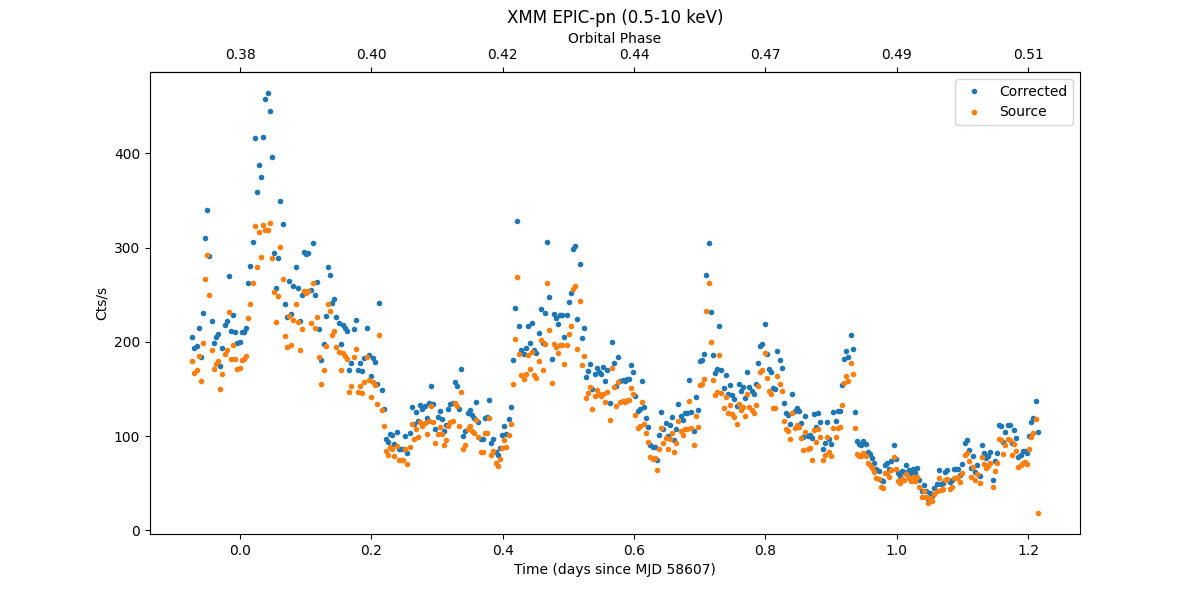}
    \caption{Light curve of EPIC-pn with 283 s bins. Short flares can be observed as sharp edges, with a longer flaring episode of $\approx 8\,\mathrm{ks}$ from Time $ > 0.4$ days since MJD 58607 (May 4th 12AM UTC). Orbital phase calculated with $T_{90}$ with ephemeris from \cite{Kreykenbohm_2008}.}
    \label{fig:lightcurvetimeresolution283}
\end{figure}

\newpage

\subsection{Scripts Used for Science Case Study} \label{app:scriptsforvela}
To achieve the pySAS extraction of the science case study in sec.\ref{sec:case_study}, we have utilised multiple Python scripts. These scripts can be found available at the GitHub repository: \url{https://github.com/esa/xmm-sas-datalabs-case-study} for public use and reproducibility of our results. The scripts are written to complement the Jupyter Notebook \texttt{CaseStudy.ipynb} also available in the before-mentioned repository.

\begin{table}[h]
    \centering
    \begin{tabular}{|c|p{10cm}|}
        \hline
        \textit{Python file} &  \textit{Description} \\
        \hline
        \texttt{plotVelaX1LC} (\textit{function in} \texttt{plotLC.py}) & Plots the light curve of Vela X-1 using Matplotlib. It extracts time and count rate data from FITS files, converts time to days since MJD 58607, and overlays orbital phase information. It supports optional logarithmic scaling, threshold lines, and point connection. See outputs of this function in fig.\ref{fig:lightcurvetimeresolution283} and fig.\ref{fig:timeresolvedLC}.\\
        \hline
        \texttt{energy-resolvedLC.py}  &  This script extracts light curves for the each four bands of interest for further analysis of variability in photon count rates. The final energy-resolved light curves are saved in a directory and appended to a list for further analysis. This script is used to produce the light curves presented in fig.\ref{fig:timeresolvedLC}.\\
        \hline
        \texttt{spectrum-extractor.py}  & This script calculates time intervals for spectral extraction from three orbital phase points, sets up the SAS environment, and iteratively extracts source spectra from specific detector regions based on time filtering. The script then generates response (RMF) and ancillary (ARF) files, applies spectral grouping, and outputs the final grouped spectra. This script is used to produce spectra presented in fig.\ref{fig:3phasesepctra}.\\
        \hline
        \texttt{xspecplot.py} &  Generates stacked spectral plots using XSPEC and Matplotlib. It retrieves spectral data (energy values, count rates, and errors) from XSPEC, applies logarithmic scaling if specified, and overlays reference lines at specific energies. The function then customizes the plot appearance and saves the figure. See output of this script in fig.\ref{fig:3phasesepctra}. \\
        \hline
        \texttt{gtiloop.py}  & This script initializes the SAS environment and reads the event file to extract the observation start and end times. It then creates Good Time Interval (GTI) files by iterating over the pulse period in 283.44-second intervals using the \texttt{tabgtigen} command. The resulting GTI files are saved in a specified directory.\\
        \hline
        \texttt{loopgtispectra.py} & This script iterates over GTI files, produced running \texttt{gtiloop.py}, to extract spectra while avoiding pile-up regions using \texttt{evselect}, then applies background scaling (\texttt{backscale}), response matrix generation (\texttt{rmfgen}), and ancillary response file creation (\texttt{arfgen}). Finally, it groups the spectra using \texttt{specgroup} and saves the outputs.\\
        \hline
    \end{tabular}
    \caption{Description of Python scripts used and written for the swifter pySAS extraction of Vela X-1 at sec.\ref{sec:case_study}. We have also used some of the already provided Python tools available in the Datalab such as \texttt{js9helper.py} and \texttt{plotLC.py}, where the latter was modified to include an extra function. See table \ref{tab:pyfunc} for descriptions.}
    \label{tab:pyfuncvela}
\end{table}

\twocolumn
\bibliographystyle{elsarticle-harv} 
\bibliography{sas_datalabs.bib}






\end{document}